\documentclass[prb,amssymb,amsmath,twocolumn,showpacs]{revtex4}
\usepackage{bm}
\usepackage{graphicx}
\usepackage{color}

\begin{document}

\title{Transmission coefficient through a saddle-point
electrostatic potential for graphene in the quantum Hall regime}

\author{Martina Fl\"{o}ser}
\affiliation{Institut N\'{e}el, CNRS and Universit\'{e} Joseph Fourier, B.P.
166, 25 Avenue des Martyrs, 38042 Grenoble Cedex 9, France}

\author{Thierry Champel}
\affiliation{Laboratoire de Physique et Mod\'{e}lisation des Milieux
Condens\'{e}s, CNRS and Universit\'{e} Joseph Fourier, B.P. 166, 25 Avenue des
Martyrs, 38042 Grenoble Cedex 9, France}

\author{Serge Florens}
\affiliation{Institut N\'{e}el, CNRS and Universit\'{e} Joseph Fourier, B.P.
166, 25 Avenue des Martyrs, 38042 Grenoble Cedex 9, France}

\date{\today}

\begin{abstract}
 From the scattering of semicoherent-state wave packets at high magnetic field,
we derive analytically the transmission coefficient of electrons in graphene 
in the quantum Hall regime through a smooth constriction described by a quadratic
saddle-point electrostatic potential.
We find anomalous half-quantized conductance steps that are rounded by a backscattering
amplitude related to the curvature of the potential. Furthermore, the conductance
in graphene breaks particle-hole symmetry in cases where the saddle-point potential is
itself asymmetric in space.
These results have implications both for the interpretation of split-gate transport 
experiments, and for the derivation of quantum percolation models for graphene.
\end{abstract}


\maketitle


The quantum point-contact geometry formed by metallic split gates in the quantum
Hall regime is a cornerstone of many experiments in two-dimensional electron
gases (2DEGs) based on semiconducting heterostructures. For example, in recent
interferometry experiments made with electrons, the quantum point contact plays
the role of an electronic beam splitter. \cite{Neder2007} At the
theoretical side, the consideration of a smooth constriction represents a simple
toy model of elaborated quantum transport theories. For instance, the
transmission coefficient through a saddle-point potential \cite{Fertig1987} is a
central piece of the percolation network models, \cite{Chalker1988,Kramer2005}
which have been introduced to describe the inter-plateaus dissipative transport
in the quantum Hall effect.

Smooth constrictions are usually modeled by the local potential profile at the
bottleneck of the constriction
\begin{eqnarray}
V({\bf r})=by^{2} - a x^{2}
,
\label{pot}
\end{eqnarray}
where $a$ and $b$ are real positive coefficients characterizing the potential
(for convenience, we chose the electrostatic potential value at the saddle
as the origin for the energies). Important insights on the tunneling processes
determining the transport properties through quantum point contacts can be
gained from the quantum mechanical motion of the electron in such a simple potential as given by Eq. \eqref{pot}.
In the standard 2DEGs usually described by Schr\"{o}dinger's equation, the energy
dependence of the transmission coefficient through the quadratic potential in Eq.
\eqref{pot} is well-known and given under high magnetic fields (i.e., neglecting
Landau-level mixing while keeping the magnetic length $l_{B}$ finite) by the
expression \cite{note1}
\begin{eqnarray}
T_{n}(E)= \left[
1
+
 \exp\left(-\pi \frac{E-\{n+1/2\}(\hbar \omega_{c}+\zeta)}{l_{B}^{2} \sqrt{ab}} \right)
\right]^{-1}
\label{T2DEG}
\end{eqnarray}
for the $n$th Landau level, where $l_{B}=\sqrt{\hbar c/(|e|B)}$ is the
magnetic length, $\hbar \omega_{c}$ is the Landau-level spacing in the 2DEG, and
$\zeta=l_{B}^{2}(b-a)$. Tunneling processes give rise to a nonzero probability
for the electron at energy $E <0$ to be transmitted on the other side of
the constriction while an electron at $E >0$ goes weakly backscattered through 
the available channels.

With monolayer graphene consisting of carbon atoms packed in a two-dimensional
honeycomb lattice appears a new host material, \cite{Geim2007} where electrons
are confined to two dimensions, yet with some exotic properties. The observation
of an anomalous quantum Hall effect in graphene \cite{Novo2005,Zhang2005}
understood in terms of a relativisticlike spectrum of low-energy electrons
\cite{Gus2005} has been followed by numerous experimental and theoretical
contributions \cite{Castro2009} aiming at exhibiting specific signatures of the
2D massless Dirac fermions in a non-uniform potential. The studies of simple
analytical problems, such as the quantum-mechanical motion of massless particles
in the quadratic saddle-point potential [\eqref{pot}], are of valuable interest
for identifying such unusual properties. However, it turns out that quadratic
potentials, which are exactly solvable in ordinary 2DEGs at any magnetic field,
become generally not analytically solvable within the Dirac equation. Here, we
consider the regime of large magnetic fields, in which the Landau level
spacing in graphene is large enough so that one can work in the single Landau-level limit.
We have recently shown \cite{Champel2010} using a semicoherent-state Green's-function
formalism that quadratic problems then become soluble in this regime.


The determination of the transmission coefficient through a smooth constriction
for graphene in the high magnetic field limit [i.e., the counterpart of Eq. (\ref{T2DEG})] 
is the main result of this Rapid Communication, that we start by discussing in relation to conductance quantization 
in graphene, with a detailed derivation using semicoherent-state Green's functions making 
the rest of the Rapid Communication.
The exact transmission coefficient for graphene in the absence of Landau-level mixing 
reads in the $n$th Landau level (with $n \geq 1$)
\begin{eqnarray}
T_{n,\epsilon}(E)= \left[
1
+
 \exp\left(-\epsilon \pi \frac{E-E_{n,\epsilon}}{ l_{B}^{2} \sqrt{ab}} \right)
\right]^{-1}, \label{resultT}
\end{eqnarray}
where $\epsilon=\pm 1$ is a band index characterizing the electron and hole-like contributions, and
\begin{eqnarray}
E_{n,\epsilon}=n \zeta +\epsilon \sqrt{n \left(\hbar \Omega_{c} \right)^{2}
+\zeta^{2}/4} \label{En}
\end{eqnarray}
with $\Omega_{c}= \sqrt{2}v_{F}/l_{B}$ ($v_{F}$ is the graphene Fermi velocity).
The transmission probability for the lowest Landau level ($n=0$) is given by
$T_{0}(E)=1/2$.

The zero-temperature conductance at chemical potential $\mu$ is given in
terms of the transmission probabilities by the Landauer-B\"{u}ttiker formula
\begin{eqnarray}
{\cal G}(\mu) = \frac{4 e^{2}}{h} 
 \left[ T_{0}(\mu) 
+\sum_{n=1}^{+\infty} \sum_{\epsilon=\pm} T_{n,\epsilon}(\mu) \right],
\end{eqnarray}
where we have accounted for the spin and valley degeneracies in graphene with the 
overall prefactor 4.

\begin{figure}[ht]
\includegraphics[scale=0.6]{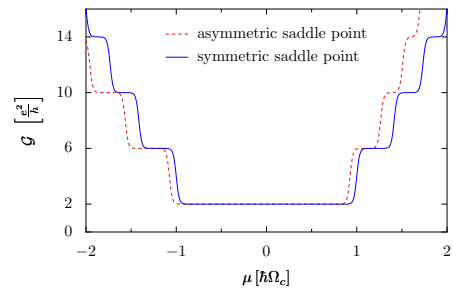}
\caption{(Color online) Zero-temperature conductance for a saddle-point
electrostatic potential with $l_B^2 a= l_B^2 b=0.05\Omega_c$ (symmetric case) 
and $l_B^2 a=4l_B^2 b=0.1\Omega_c$ (asymmetric case).
Asymmetries of the electrostatic saddle-point potential reflect in 
asymmetries between the electron and hole sectors for the conductance.}
\label{Conductance}
\end{figure}

Then, as found in Ref. \onlinecite{Buttiker1990} for the 2DEG case, the conductance
quantization in the graphene case shown in Fig. \ref{Conductance} directly
follows from the transmission probabilities. The first obvious observation is
the half-integer quantization of the conductance in terms of the conductance
quantum (here $4 e^{2}/h$) with plateaus at values $(n+1/2)4e^{2}/h$,
reminiscent of the half-integer quantization of the Hall conductance. Two
different configurations, symmetric and asymmetric with respect to $\pi/2$
rotation of the saddle-point potential in Eq. \eqref{pot}, have been considered, which yield 
to two different curves.
A symmetric saddle-point potential is characterized by $a=b$, thus $\zeta=0$,
and the conductance shows clear particle-hole symmetry with respect to
change in the energy sign. In the case of an asymmetric saddle-point potential ($a \neq b$), $\zeta$ becomes non-zero, signaling a breaking of
particle-hole symmetry in the energy levels~Eq. \eqref{En}, and resulting in a 
non-uniform shift of the conductance steps, which is more pronounced for the 
highest Landau levels.
It is worth noting that the asymmetry of the potential in Eq. \eqref{pot} has a different 
consequence in the case of the standard 2DEG, where it just leads to a redefinition
of the Landau-level spacing, see Eq. (\ref{T2DEG}), and such a small quantitative 
modification appears difficult to perceive in an experiment. 
In graphene, the effect of the potential asymmetry should be more easily seen in 
experiments, since it yields an asymmetry between the positive- and negative- energy 
dependences of the conductance. 
%

Now, we focus on the derivation of Eq. (\ref{resultT}). We consider a
single-particle Hamiltonian model for an electron of charge $e=-|e|$ and of
Fermi velocity $v_{F}$ confined to a two-dimensional graphene sheet in the plane
$(x,y)$ in the presence of both a perpendicular uniform magnetic field ${\bf
B}=B \hat{{\bf z}}$ and an electrostatic (scalar) potential term $V({\bf r})$ 
[given by Eq. \eqref{pot}]
\begin{eqnarray}
\hat{H}=v_{F} \left( \begin{array}{cc}
0 & \Pi_{x}-i\Pi_{y} \\
\Pi_{x}+i\Pi_{y} & 0
\end{array}
\right)+V({\bf r}) \hat{1}
\label{Ham}
\end{eqnarray}
with ${\bf \Pi}=-i \hbar {\bm \nabla}_{{\bf r}}-e {\bf A}({\bf r})/c$, where
${\bf A}({\bf r})$ is the vector potential defined by the equation ${\bm
\nabla}_{\bf r} \times {\bf A}({\bf r}) ={\bf B}$. Here $\hat{1}$ corresponds
to the unity matrix in the pseudo-spin space (representing electron and hole
degrees of freedom) and $c$ is the speed of light. For convenience, we will
omit both physical spin and valley indices, thus assuming that the two valleys
of graphene remain completely decoupled form each other and can be studied
separately. In addition, we shall not consider the effects of ripples or a mass
potential but these could be studied following Ref.~\onlinecite{Champel2010}.

To describe the electron dynamics at high magnetic field, it is
useful\cite{Champel2010} to introduce the graphene vortex states 
\begin{eqnarray}
\tilde{\Psi}_{n,{\bf R},\lambda}({\bf r})
&=&\frac{1}{\sqrt{1+|\lambda|}} \left(
\begin{array}{c}
\lambda \Psi_{n-1,{\bf R}}({\bf r}) \\
i \Psi_{n,{\bf R}}({\bf r})
\end{array}
\right), \label{vort} \\
\Psi_{n,{\bf R}}({\bf r})
&=&
 \frac{e^{-\left(|z|^{2}+|Z|^{2}-2Zz^{\ast} \right)/(4l_{B}^{2})}
}{\sqrt{2 \pi l_{B}^{2}n!}}
\left(\frac{z-Z}{\sqrt{2} l_{B}} \right)^{n}
\end{eqnarray}
with $z=x+iy$ and $Z=X+iY$. Here ${\bf R}=(X,Y)$ is a doubly continuous quantum
number corresponding to the guiding center position in the plane, $n$ is a
positive integer, $\lambda$ is a band index (defined for a given $n$), which is
equal to $\pm 1$ if $n \geq 1$, and 0 for $n=0$. States in Eq. \eqref{vort}, which can
be written as $\tilde{\Psi}_{n,{\bf R},\lambda}({\bf r}) =\langle {\bf r} |
n,{\bf R},\lambda \rangle$ within the Dirac bracket notation, are eigenstates
of Hamiltonian (\ref{Ham}) in absence of the electrostatic potential ($V=0$) with
the energy quantization $E_{n,\lambda}=\lambda \sqrt{n} \hbar \Omega_{c}$.
Despite being nonorthogonal with respect to the degeneracy quantum number 
${\bf R}$
\begin{eqnarray}
\langle n_{1},{\bf R}_{1},\lambda_{1} | n_{2},{\bf R}_{2},\lambda_{2}\rangle
=
\delta_{n_{1},n_{2}} \langle {\bf R}_{1} |{\bf R}_{2} \rangle \delta_{\lambda_{1},\lambda_{2}} 
, \hspace*{1.5cm} \\
 \langle {\bf R}_{1} |{\bf R}_{2} \rangle =
\exp \left[- \frac{({\bf R}_{1}-{\bf R}_{2})^{2} - 2 i \hat{{\bf z}} \cdot ({\bf R}_{1} 
\times {\bf R}_{2})}{4 l_{B}^{2}} \right]
\end{eqnarray}
the set of quantum numbers $|n,{\bf R},\lambda \rangle$ obeys a completeness relation
\begin{eqnarray}
\int \frac{d^{2} {\bf R}}{2 \pi l_{B}^{2}} \sum_{n=0}^{+\infty} \sum_{\lambda} |n,{\bf R},
\lambda \rangle \langle n,{\bf R},\lambda | = \hat{1}. \label{comp}
\end{eqnarray}
In fact, the states in Eq. \eqref{vort} form an overcomplete basis of states, which have
the {\it coherent states character} with respect to the quantum number ${\bf R}$.

Relation (\ref{comp}) allows one to project the electron dynamics onto the
vortex representation. We can then introduce the vortex Green's function
$G(n_{1},{\bf R}_{1},\lambda_{1},t_{1} ; n_{2},{\bf R}_{2},\lambda_{2},t_{2})$,
which gives the probability amplitude for a vortex with circulation (or
Landau-level index) $n_{1}$ and band index $\lambda_{1}$ that is initially at
position ${\bf R}_{1}$ at time $t_{1}$ to be at point ${\bf R}_{2}$ at time
$t_{2}$ with a new circulation $n_{2}$ and a band index $\lambda_{2}$. In the
following, we consider the dynamics projected onto a single Landau level,
meaning that the vortex circulation is conserved ($n_{1}=n_{2}=n$). Formally,
this corresponds to taking the limit $v_{F} \to + \infty$. 
Within a single Landau level $n$, the retarded Green's function takes the 
form \cite{Champel2010} 
\begin{eqnarray}
G_{n;\lambda_{1};\lambda_{2}}({\bf R}_{1},{\bf R}_{2})
= \langle {\bf R}_{1} | {\bf R}_{2} \rangle e^{(l_{B}^{2}/4)\Delta_{{\bf R}_{12}} } \left[ 
\tilde{g}_{n;\lambda_{1};\lambda_{2}}({\bf R}_{12} ) \right] \label{G},
\\
{\bf R}_{12}=\frac{1}{2} \left[ 
{\bf R}_{1}+{\bf R}_{2}+i ({\bf R}_{2}-{\bf R}_{1}) \times \hat{{\bf z}}
\right]
, \hspace*{1cm}
\end{eqnarray}
where $\Delta_{{\bf R}}$ is the Laplacian operator. In the absence of
Landau-level mixing and for a quadratic saddle-point electrostatic potential with the
spatial dependence given by Eq. \eqref{pot}, the function $\tilde{g}$ can be calculated
exactly \cite{Champel2010} with the result in the energy representation (i.e.,
after Fourier transformation with respect to the time difference
$t_{1}-t_{2}=t$)
\begin{eqnarray}
\tilde{g}_{n;\lambda_{1};\lambda_{2}}({\bf R})
=
 \int_{0}^{+ \infty} \!\!\! \!\!\! dt \, \frac{-i
e^{-i \tau(t) V({\bf R})} 
}
{ \cosh\left(\sqrt{|\gamma|}t\right)}
h_{n;\lambda_{1};\lambda_{2}}(t) \,
 e^{it(E+i0^{+})}
,
\label{gn}
\end{eqnarray}
where $\tau(t)=(1/\sqrt{|\gamma|})\tanh (\sqrt{|\gamma|}t)$.
Here $E$ is the energy and $0^{+}$ an infinitesimal positive quantity.
The parameters $\gamma$ and $\zeta$ are geometric coefficients characterizing $V({\bf r})$ 
\begin{eqnarray}
\gamma = \frac{l_{B}^{4}}{4} \left[ 
\left( \partial^{2}_{x}V \right) \left(\partial^{2}_{y}V\right)-\left( \partial_{x}\partial_{y}V\right)^{2}
\right], \hspace*{0.1cm}
\zeta =\frac{l_{B}^{2}}{2} \Delta_{{\bf r}} V({\bf r}).
\end{eqnarray}
The coefficient $\gamma$ is directly proportional to the Gaussian curvature of
the electrostatic potential, which turns out to be negative for a saddle-shaped (or
hyperbolic) quadratic function $V({\bf r})$. Finally, the functions
$h_{n;\lambda_{1};\lambda_{2}}(t)$, which contain the full dependences on the
Landau-level index $n$ and on the band indices $\lambda_{1}$ and $\lambda_{2}$
read for $n \geq 1$
\begin{eqnarray}
h_{n;\lambda_{1};\lambda_{2}}(t)
= 
\sum_{\epsilon=\pm}
\left[ 
(1+\epsilon \lambda_{1} \alpha_{n}) \, \delta_{\lambda_{1},\lambda_{2}}+
\epsilon \beta_{n} \delta_{-\lambda_{1},\lambda_{2}}
\right] \frac{
 e^{-it E_{n,\epsilon}}
}{2}, \hspace*{-1cm}
\nonumber \\
 \label{hn}
\end{eqnarray}
where $E_{n,\epsilon}$ is defined in Eq. (\ref{En}),
\begin{eqnarray}
\alpha_{n} = \frac{\sqrt{n} \hbar \Omega_{c}}{\sqrt{n \left(\hbar \Omega_{c} \right)^{2}
+\zeta^{2}/4}}, \hspace*{0.5cm}
\beta_{n} = \frac{\zeta}{ \sqrt{n \left(\hbar \Omega_{c} \right)^{2}
+\zeta^{2}/4}}
,
\end{eqnarray}
and $h_{0;0;0}(t)=e^{-it \zeta/2}$ for the lowest Landau level $n=0$. It is worth
stressing that the previous expressions are valid for Landau-level indices $n$
not too high, for which the inequality $|\zeta|, \sqrt{|\gamma|} \ll
(\sqrt{n+1}-\sqrt{n}) \hbar \Omega_{c}$ holds.

The action of the differential operator $\exp \left[ (l_{B}^{2}/4) \Delta_{{\bf
R}} \right]$ on the function $\tilde{g}({\bf R})$ in Eq. (\ref{G}) is evaluated as
\cite{Champel2009}
\begin{eqnarray}
G_{n;\lambda_{1};\lambda_{2}}({\bf R}_{1},{\bf R}_{2})
= \langle {\bf R}_{1} | {\bf R}_{2} \rangle \!\! 
\int \!\! \frac{d^{2} {\bf u}}{\pi l_{B}^{2}} \tilde{g}_{n;\lambda_{1};\lambda_{2}}({\bf u}) \, e^{- \frac{({\bf u}-{\bf R}_{12})^{2}}{l_{B}^{2}} }. \nonumber \hspace*{-0.5cm} \\
\label{conv}
\end{eqnarray}
Then, inserting Eq. (\ref{gn}) into Eq. (\ref{conv}),
 we can perform the Gaussian integrals over ${\bf u}$ to get
\begin{eqnarray}
G_{n;\lambda_{1};\lambda_{2}}({\bf R}_{1},{\bf R}_{2})
= \langle {\bf R}_{1} | {\bf R}_{2} \rangle \!\! 
\int_{0}^{+ \infty} \!\!\!\! \!\!\!\! dt \,
\frac{
 -i e^{it(E+i0^{+})}
}
{\cosh\left(\sqrt{|\gamma|}t\right)} \, h_{n;\lambda_{1};\lambda_{2}}(t) 
\nonumber \hspace*{-1cm} \\
\times
\sqrt{f(t)} \, e^{-i f(t) \tau(t) V({\bf R}_{12})} \, e^{\gamma f(t) \tau^{2}(t) \frac{{\bf R}_{12}^{2} }{l_{B}^{2}}}
, \hspace*{0.5cm} \label{intG}
\end{eqnarray}
with $f(t)=\left(1+i \zeta \tau(t)-\gamma \tau^{2}(t) \right)^{-1}$.

We note from Eq. \eqref{hn} that for $n \geq 1$ and $\zeta \neq 0$ the function
$h_{n;\lambda_{1};\lambda_{2}}(t)$ is not diagonal in the $\lambda$ space,
indicating that $\lambda$ is generically no more a good quantum number in the
presence of asymmetric saddle-point potentials. A straightforward
diagonalization shows that $\epsilon=\pm$ appears instead as a good number, with
$h_{n;\epsilon}(t)=e^{-it E_{n,\epsilon}}$. According to the above formula
(\ref{En}) for $E_{n,\epsilon}$, the quantum number $\epsilon$ clearly labels
the electron-like and hole-like energy bands. We shall henceforth represent the
Green's function in this $\epsilon$ representation, where the latter takes a
diagonal form.


In order to determine the transmission coefficient with a given energy channel
(i.e., at Landau-level index $n$ and band index $\epsilon$ fixed), we only need
the Green's function $G_{n;\epsilon}({\bf R}_{1},{\bf R}_{2})$ when the states
at vortex positions ${\bf R}_{1}$ and ${\bf R}_{2}$ correspond to the same
energy and are asymptotically far from the saddle point located at the origin.
For a saddle-point potential of the form (\ref{pot}), this means taking the
limits $|X_{1}| \to \infty$ and $|X_{2}| \to \infty$, while
$V(X_{1},Y_{1})=V(X_{2},Y_{2})=const$. Making the change in variable 
$
s= d \left[1-|\gamma|\tau^{2}(t) \right]/\left[1+|\gamma| \tau^{2}(t) \right]$,
where $d=|X_{1}Y_{2}-X_{2}Y_{1}|/l_{B}^{2}$,
 we can easily take the limit $d \to + \infty$ in the integral in Eq. \eqref{intG}, and obtain the expression
\begin{eqnarray}
G^{\infty}_{n;\epsilon}({\bf R}_{1},{\bf R}_{2})
=
\Gamma\left(\frac{1}{2}-i \frac{E-E_{n,\epsilon}}{2 \sqrt{|\gamma|}} \right)
e^{-\sigma \frac{\pi}{4}\left(\frac{E-E_{n,\epsilon}}{\sqrt{|\gamma|}} + i\right)} 
\nonumber \\
\times (- i) 
e^{i\frac{\left(\sqrt{a}X_{1}-\sqrt{b} Y_{1} \right)\left( \sqrt{b}X_{1}+\sqrt{a}Y_{1}\right)+
\left(\sqrt{a}X_{2}+\sqrt{b} Y_{2} \right)\left( \sqrt{b}X_{2}-\sqrt{a}Y_{2}\right)
}{2 l_{B}^{2}(a+b)}} 
\nonumber \\
\times
\frac{ 
 d^{-1/2+ i \left( E-E_{n,\epsilon} \right)/(2 \sqrt{|\gamma|})}
}{\sqrt{4|\gamma|+2i \zeta \sqrt{|\gamma|}}} \, 
e^{- \frac{\left(\sqrt{a}X_{1}-\sqrt{b} Y_{1} \right)^{2}+
\left(\sqrt{a}X_{2}+\sqrt{b} Y_{2} \right)^{2}
}{2 l_{B}^{2}(a+b)}}, \nonumber \\
\hspace*{0.5cm} 
\label{Gas2}
\end{eqnarray}
where $\sigma=\mathrm{sgn}(X_{1}X_{2})$ and $\Gamma(z)$ is the Gamma function.
%
At infinity the vortex is close to the asymptotes of the saddle-point
potential $V({\bf R})$, i.e., $Y_{1} \sim X_{1} \sqrt{a/b} $ and $Y_{2} \sim
- X_{2} \sqrt{a/b} $. The two modes $\epsilon = \pm$ circulate in the same
direction, and are not mixed since they are well separated in energy.
\begin{figure}[t!]
\includegraphics[scale=0.6]{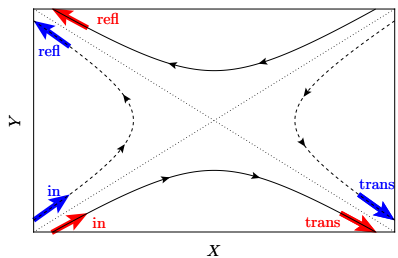}
\caption{(Color online) Schematic to identify the Green's functions for reflection
and transmission. Semiclassically the trajectories of vortex states lie on
contours of constant potential. Solid (dashed) lines correspond to an
equipotential line $V>0$ ($V<0$) and therefore for $E>0$ to a vortex in the
conduction (valence) band, respectively. Dotted lines denote the asymptotes of the 
potential in Eq. \eqref{pot}.}
\label{fig:schema}
\end{figure}

Within the present saddle-point geometry, the transmission coefficient can 
be extracted from the asymptotic form of the retarded vortex Green's function,
thus performing scattering theory in terms of coherent-state wave packets instead 
of the more standard plane waves. 
In the conduction band ($\epsilon=+$), the transmission of a vortex from the
left half-plane to the right half-plane is described by Green's function
expression [Eq. \eqref{Gas2}] with the sign function $\sigma=-$ while the vortex
reflection, where the vortex remains in the left half-plane, is characterized by
$\sigma=+$ (see Fig. \ref{fig:schema}).
Therefore, the ratio of the transmission amplitude to the reflection amplitude
for a given Landau level $n$ in the conduction band is given by the ratio of
these two Green's functions:
\begin{eqnarray}
\frac{t_{n,+}}{r_{n,+}}
= i \exp\left[\frac{\pi}{2}\frac{E-E_{n,+}}{\sqrt{|\gamma|}} \right].
\end{eqnarray}
In the valence band the vortices carry a positive charge. Therefore, the
correspondence between the transmission (or reflection) process of a negative
charge carrier and the sign of $\sigma$ is now reversed,
so that we have 
\begin{eqnarray}
\frac{r_{n,-}}{t_{n,-}}
=
i \exp\left[\frac{\pi}{2}\frac{E-E_{n,-}}{\sqrt{|\gamma|}} \right].
\end{eqnarray}
The relations between the transmission probabilities $T_{n,\epsilon}$ and
the transmission and reflection amplitudes, i.e., 
$\left|t_{n,\epsilon}\right|^{2} = 1-\left| r_{n,\epsilon} \right|^{2}= 
T_{n,\epsilon}$, finally provide the result in Eq. \eqref{resultT}.

For the peculiar case of the lowest Landau level, contributions both from 
the original conduction and valence bands arise.
We get for the electron-like excitations 
$ T_{0}^{+}(E)= \left[ 1 +
 \exp\left(-\pi (E-\zeta/2)/\sqrt{|\gamma|} \right)
\right]^{-1}$,
and for the hole-like excitations 
$T_{0}^{-}(E)= \left[
1 + \exp\left(\pi (E-\zeta/2)/\sqrt{|\gamma|} \right)
\right]^{-1}$.
Summing up these two contributions and considering the equipartition of
the current between the two types of excitations yielding a $1/2$ prefactor, we
finally get the already quoted lowest Landau-level contribution
\begin{eqnarray}
T_{0}(E)= \frac{1}{2} \left[ T_{0}^{+}(E)+T_{0}^{-}(E) \right] =\frac{1}{2}.
\end{eqnarray}

In conclusion, we have calculated the transmission coefficient for a quadratic
saddle-point electrostatic potential in graphene, and found that shape
asymmetries generic to quantum point contacts break particle-hole symmetry
in the conductance. Our results should be relevant for future split
gate experiments in graphene as well as for the formulation of quantum network
models. Theoretically, we have also presented an alternative way of
deriving transmission coefficients from the scattering of
coherent-state wave packets. 

\end{document}